\documentclass{webofc}
\usepackage[varg]{txfonts}   

\usepackage{xspace}	
\usepackage{color}	

\newcommand{\pt}{\mbox{$p_{\rm T}$}\xspace}

\newcommand{\raa}{\mbox{$R_{\rm AA}$}\xspace}

\newcommand{\rda}{\mbox{$R_{d {\rm A}}$}\xspace}
\newcommand{\rcp}{\mbox{$R_{\rm cp}$}\xspace}
\newcommand{\Npart}{\mbox{$N_{\rm part}$}\xspace}
\newcommand{\Ncoll}{\mbox{$N_{\rm coll}$}\xspace}

\newcommand{\dNdeta}{\mbox{$dN_{\rm ch}/d\eta$}\xspace}

\newcommand{\snn}{\mbox{$\sqrt{s_{_{NN}}}$}\xspace}

\newcommand{\piz}{\mbox{$\pi^0$}\xspace}

\newcommand{\gevc}{\mbox{GeV/$c$}\xspace}

\begin{document}
\title{Systematic study of high $p_{\rm T}$ hadron production
in small collision systems by the PHENIX experiment at RHIC}
%
%

\author{\firstname{Takao} \lastname{Sakaguchi}\inst{1}\fnsep\thanks{\email{takao@bnl.gov}}, 
        \firstname{for the} \lastname{PHENIX Collaboration}}

\institute{Brookhaven National Laboratory, Physics Department, Upton, NY 11973-5000, USA.}

\abstract{%
High \pt hadrons in the small systems of $d/^3$He+Au and $p$+Al/Au
collisions have been measured at the midrapidity ($|\eta|<0.35$)
as well as
at the forward and backward rapidities ($1.2<|\eta|<2.4$) at
\snn=200\,GeV. A clear system and centrality ordering in the
$R_{p/d/{\rm HeA}}$ was observed for midrapidity \piz production.
Hadron \rcp as a function of rapidity in forward
and backward regions show an opposite trend to what HIJING++
predicted. Together with the difference of \rcp in $p$+Al/Au
collisions, the observation can be understood consistently
with the measured long-range two-particle correlations, hinting
the possible medium creation similar to A+A collisions.}
\maketitle
\section{Introduction}
\label{intro}
The small collision systems such as $p/d$+A collisions
have been considered a good laboratory to quantify cold
nuclear matter effects, a necessary baseline for
understanding the effects of the hot and dense medium
produced in A+A collisions. This assumption has been
confirmed by several measurements such as high transverse
momentum (\pt) hadrons, jets, and direct photons, in the
minimum bias $d$+Au collisions at
\snn=200\,GeV~\cite{Adler:2006wg,Adare:2012vn,Adare:2015gla};
the nuclear modification factors,
$\rda\equiv(1/T_{d{\rm A}})(dN^{d{\rm A}}/d\pt d\eta)/(d\sigma^{pp}/d\pt d\eta)$,
where the $T_{d{\rm A}}$ is the nuclear thickness function,
are consistent with unity up to high \pt, and the slight
deviation from the unity is consistent with the expectation
by parton distribution functions such as EPS09~\cite{Eskola:2009uj}.
The observation of the ridge-like
structure in the long-range azimuthal correlations
in the central $p$+Pb collisions at \snn=5.02 TeV at the
LHC~\cite{CMS:2012qk,Abelev:2012ola,Aad:2012gla},
however, called into question the view of such systems
as consisting merely of cold nuclear matter. The
study at the LHC was followed by the PHENIX experiment
at RHIC, and a finite $v_2$ of hadrons in 0–5\,\% central
$d$+Au collisions using both the two-particle angular
correlation method and the event-plane method were
observed~\cite{Adare:2013piz,Adare:2014keg,Adamczyk:2015xjc}.
These observations led the community to explore any
phenomena found in A+A collisions, in $p/d$+A collisions.

The energy loss of hard scattered partons produced in
the initial stage of the collisions, so-called jet quenching,
has been one of the key observations in confirming
the production of the QGP. The first evidence of this
phenomena was found in the yield suppression of high
transverse momentum (\pt) hadrons, the fragments of
the hard scattered partons. The measurement of high
\pt identified hadrons have been improved over the last
decade, and reached to the level that a precise quantitative
comparison of the data and theoretical models became
realized~\cite{Adare:2008cg, Adare:2012wg}. A recent study
also found that the energy loss scales with the particle
multiplicity (\dNdeta)~\cite{Adare:2015cua}. Obviously,
the high \pt hadrons will be a
powerful tool to investigate the parton degree of freedom
in the small systems like $p/d$+A collisions as well.
In addition, a systematic study of the high \pt hadron
spectra from small to large collision systems will be able
to explore the onset of QGP as a function of the collision
systems. In this paper, we show a new systematic
measurement of the high \pt \piz in $p/d/^3$He+Au
collisions at midrapidity ($|\eta|<$0.35) as well as
the high \pt hadrons in $p$+Al/Au collisions at forward
and backward rapidities ($1.2<|\eta|<2.4$), and discuss the
hint of possible hot medium creation in connection with
the long-range azimuthal correlation observed.

\section{Measurement Method}
\label{sec-1}
\begin{figure}[htbp]
\centering
\includegraphics[width=0.95\linewidth,clip]{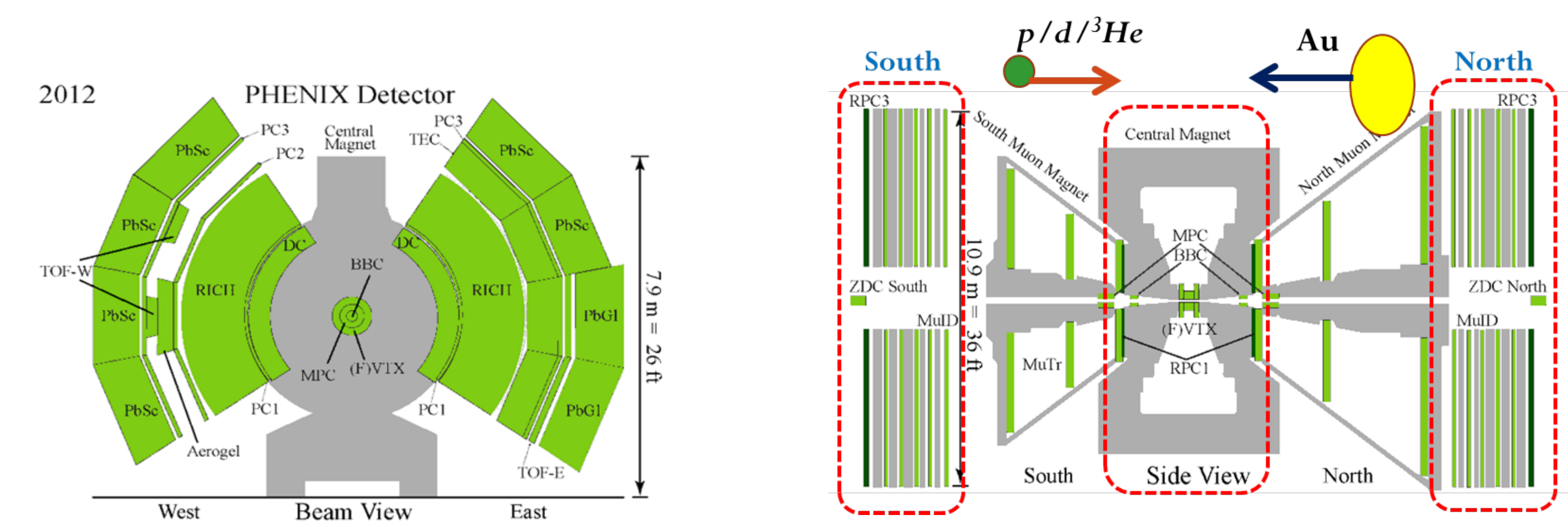}
\caption{(Left) Beam view of the PHENIX detector in the RHIC Year-2012 run and later. (Right) Side view of the PHENIX detector with the beam direction of$p/d/^3$He and Au ions shown. The central and muon arms used for this analysis are shown in the squared dotted lines. Note that Al ions go to the same directions as Au ions.}
\label{figPhenix2012}       
\end{figure}

The detector setup was the same
as the one in the RHIC Year-2012 run as shown in
Fig.~\ref{figPhenix2012}.
The detailed description of the PHENIX detector
system can be found elsewhere~\cite{Adcox:2003zm}.
The \piz's were
reconstructed via $\piz\rightarrow\gamma\gamma$,
by primarily using a lead-scintillator
sandwich type electromagnetic calorimeter
(PbSc EMCal) in the midrapidity ($|\eta|<0.35$). The threshold
of cluster energy is set to 0.2\,GeV and the photon clusters
were selected using
a shower shape cut. Then, an energy asymmetry cut of
$\alpha=|E_1-E_2|/(E_1+E_2)<0.8$ was applied on selecting
pairs of photons from \piz decay. The efficiency and
acceptance of \piz's were estimated using a GEANT based
detector simulation software.
The hadrons in the forward and backward rapidity
($1.2<|\eta|<2.4$) were measured by a muon tracker
(MuTr), a silicon vertex detector (FVTX), and a muon
identification detector (MuID). The tracks that left
signals in MuTr, FVTX and the first two sensitive
layers of MuID are identified as high \pt hadrons.

This analysis used the events recorded by PHENIX in
RHIC Year-14 ($^3$He+Au) and Year-15 ($p$+Al/Au) runs.
We used the $d$+Au results from the literature published
before~\cite{Adler:2006wg}.
The integrated luminosities recorded are
25\,nb$^{-1}$ (15\,pb$^{-1}$ pp-equivalent) for
$^3$He+Au collisions, 275\,nb$^{-1}$ (7.4\,pb$^{-1}$
pp-equivalent) for $p$+Al collisions, and 80\,nb$^{-1}$
(16\,pb$^{-1}$ pp-equivalent) for $p$+Au collisions.
Three types of events were used in this analysis; the first type of
events were triggered by the coincidence of signals from
the two Beam-Beam counters (BBC) located at $3.1<|\eta|<3.9$
covering the full azimuth (minimum bias triggered events),
the second type of events were triggered by a high energy
tower hit in the EMCal coincided with a minimum bias
trigger (ERT triggered events), and the third type of
events were triggered by a high momentum track identified
by a MuTr and MuID coincided with a minimum bias trigger
(Muon-arm triggered events).
The minimum bias trigger is not 100\,\% efficient to
the inelastic collisions
because of the limited acceptance and efficiency of the
BBC. This inefficiency increases as the collision system
becomes smaller. They were already studied in $d$+Au
collisions by comparing the charge of the south BBC
with a Glauber Monte Carlo simulation folded with a
negative binomial
distribution~\cite{Adare:2013nff}. From this comparison,
we determined that the trigger efficiency is 88\,\% for
d+Au collisions.
We followed the same method, and determined the trigger
efficiency of $p$+Al, $p$+Au and $^3$He+Au collisions also as
74\,\%, 84\,\% and 88\,\%, respectively.
In case of 200\,GeV Au+Au collisions, the efficiency
was 94\,\%~\cite{Adare:2012wg}. When dividing the events into centralities,
an additional bias factor plays a role. The bias is
originated from a correlation of particle yield
in the rapidity range for the spectra measurements,
and that in the rapidity range for the centrality measurement
which is performed by the south BBC (-3.9$<|\eta|<$-3.1).
We estimated the bias factors for the $^3$He+Au and
$p$+Al/Au collisions by following the method applied for
$d$+Au collisions~\cite{Adare:2013nff}, and determined
as listed in Table~\ref{tabBF}.
\begin{table}[htbp]
\centering
\caption{Bias factors for centrality selections in small collision systems.}
\label{tabBF}       
\begin{tabular}{lcccccccccc}
\hline\hline
Cent (\%) & 0--5 & 5--10 & 10--20 & 0--20 & 20--40 & 40--60 & 40--72 & 60--84 & 60--88 & 0--100 \\ \hline\hline
$p$+Al & 0.75 & 0.81 & 0.84 & - & 0.90 & - & 1.04 & - & - & - \\
$p$+Au & 0.86 & 0.90 & 0.94 & 0.90 & 0.98 & 1.02 & - & 1.00 & - & 0.86 \\
$d$+Au & 0.86 & 0.90 & 0.94 & 0.90 & 0.98 & 1.02 & - & - & 1.00 & 0.86 \\
$^3$He+Au & 0.86 & 0.90 & 0.94 & 0.90 & 0.98 & 1.02 & - & - & 1.00 & 0.86 \\
\hline\hline
\end{tabular}
\end{table}
All the results (including minimum bias and centrality-dependent)
shown in this paper are divided by these numbers after being
corrected for acceptances and efficiencies. The numbers for 0--100\,\%
will be used to obtain 0--100\,\% centrality result from the minimum bias
data of the given collision system. For instance, in order to obtain
0--100\,\% result for $d$+Au collisions, the spectra for minimum bias
events (0--88\,\% in this case) will be divided by 0.86.

\section{Results}
\label{sec-2}

\subsection{Mid-rapidity measurement}
\begin{figure}[htbp]
\centering
\includegraphics[width=1.0\linewidth,clip]{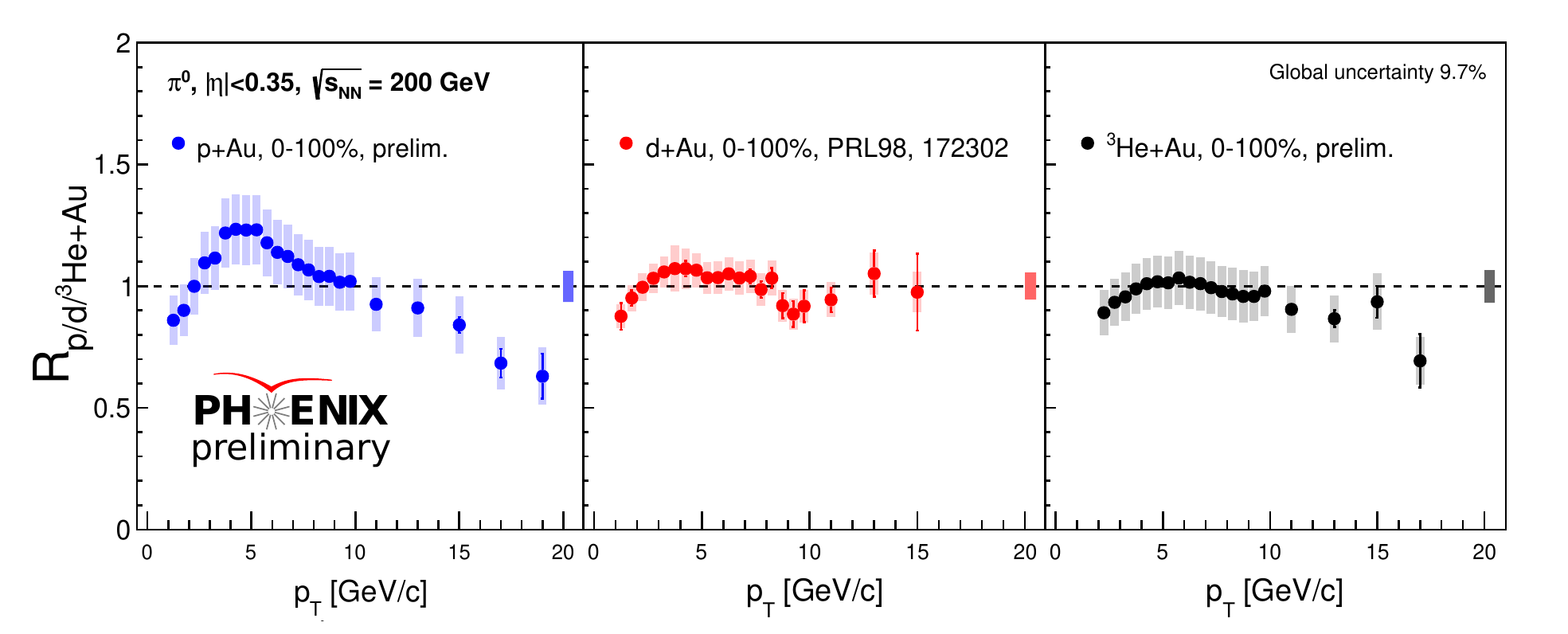}
\caption{$R_{p/d/{\rm HeA}}$ of the \piz for minimum bias $p$+Au, $d$+Au,
and $^3$He+Au collisions at \snn=200\,GeV (from left to right). Note that
the \Npart and \Ncoll are different for each system.}
\label{figRAApizMB}       
\end{figure}

\begin{figure}[htbp]
\centering
\includegraphics[width=0.95\linewidth,clip]{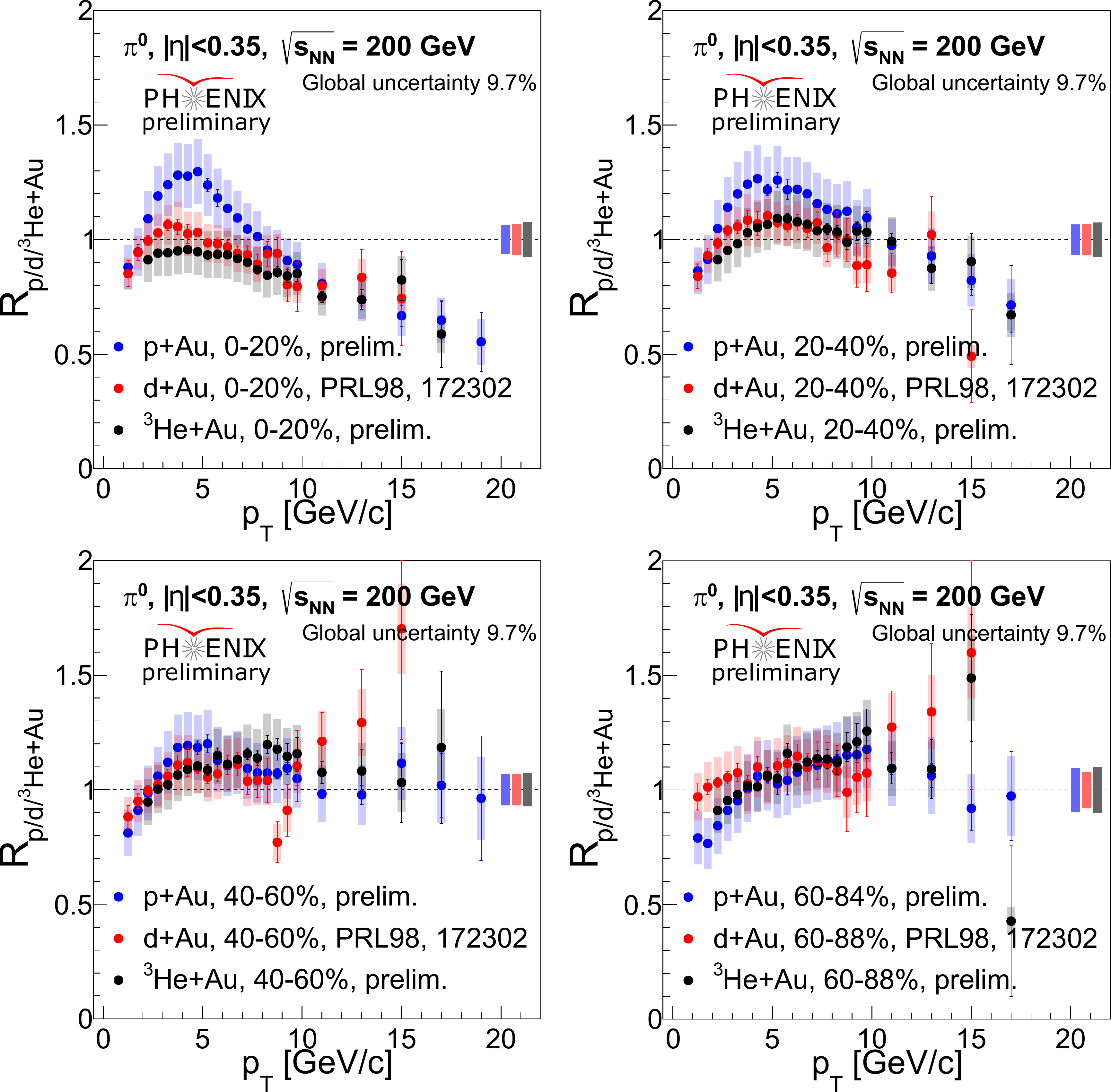}
\caption{Centrality dependence of the
\piz $R_{p/d/{\rm HeA}}$ for $p$+Au, $d$+Au, and $^3$He+Au collisions at
\snn=200\,GeV.}
\label{figRAApizcent}       
\end{figure}

\begin{figure}[htbp]
\centering
\includegraphics[width=0.95\linewidth,clip]{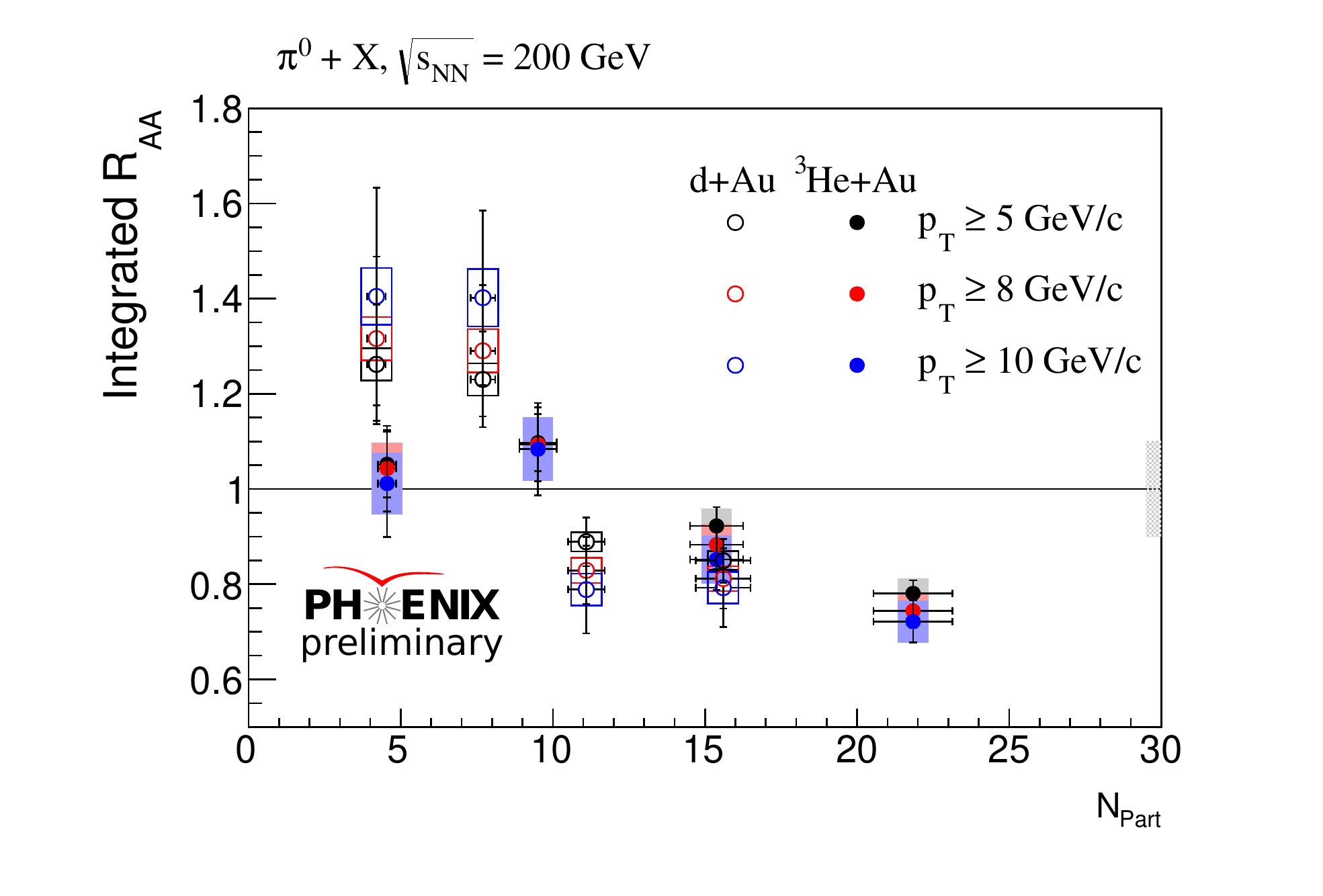}
\caption{Integrated \raa as a function of \Npart for $d$+Au and
$^3$He+Au collisions at \snn=200\,GeV.}
\label{figIntegRAA}       
\end{figure}

Figure~\ref{figRAApizMB} shows $R_{p/d/{\rm HeA}}$ of the \piz's   
for the minimum bias $p$+Au, $d$+Au, and $^3$He+Au collisions
at \snn=200\,GeV.
$R_{p/d/{\rm HeA}}$ is defined as
$R_{p/d/{\rm HeA}}\equiv(1/T_{p/d/{\rm HeA}})(dN^{p/d/{\rm HeA}}/d\pt d\eta)/(d\sigma^{pp}/d\pt d\eta)$,
where the $T_{p/d/{\rm HeA}}$ is the nuclear thickness function.
This measurement was performed at the midrapidity ($|\eta|<$0.35).
The magnitudes of the peaks seen at $\pt\sim5$\,\gevc,
which are often called as Cronin-peak and understood
as the consequence of the initial momentum broadening in the nucleus,
are found to change as a function of the collision systems. As the projectile
becomes lighter, the magnitude of the peak becomes larger.
There is a common feature over the
systems that some hint of suppression is seen at high \pt ($\pt>10$\,\gevc).
The systematic uncertainties on data points are shown in boxes on them,
the ones for $T_{p/d/{\rm HeA}}$ are
shown as the boxes around unity at the right end of each plot, and the
global uncertainty of 9.7\,\% which comes from $p+p$ luminosity
normalization is not shown but is mentioned by texts. Note that both the
number of
participant nucleons (\Npart) and the number of binary nucleon-nucleon
collisions (\Ncoll) are different for each system.
Figure~\ref{figRAApizcent} shows the centrality dependence of the
\piz $R_{p/d/{\rm HeA}}$ for $p$+Au, $d$+Au, and $^3$He+Au collisions at
\snn=200\,GeV.
The similar feature for Cronin-peaks at $\pt\sim5$\,\gevc is seen
for all the systems and centralities. The slight suppression
at higher \pt ($\pt>10$\,\gevc) tends to disappear and turns to
an enhancement as going to more peripheral collisions for all the
collision systems. The highest magnitude of
the Cronin-peak is consistently seen for $p$+Au collisions and
then $d$+Au and $^3$He+Au collisions, following the projectile
order, except for the most peripheral bin. At high \pt the
$R_{p/d/{\rm HeA}}$ of three collision systems
tend to agree each other within the quoted uncertainties for
all the centralities. Note that the same centrality doesn't mean
the \Npart and \Ncoll are same.
In order to see the development of the suppression/enhancement
at $\pt\sim$5, 8 and 10\,\gevc, we have plotted the \raa as a
function of \Npart and shown in Figure~\ref{figIntegRAA}
(only $d$+Au and $^3$He+Au collisions are shown here).
For the lower \Npart (\Npart$<\sim$10), the $R_{d/{\rm HeA}}$
are different between two systems, while for high \Npart, they
tend to merge. The result can be compared with the
\raa's from the peripheral Au+Au
collisions, i.e., 60--70, 70--80, and 80--93\,\% centrality.
In the previous publication, we measured the \raa for
\pt$>$5\,GeV/$c$ as $\sim$0.78, $\sim$ 0.87, and $\sim$0.84,
for \Npart of 26.7$\pm$3.7, 13.7$\pm$2.5, and 5.6$\pm$0.8,
respectively~\cite{Adare:2012wg}. With these numbers, we found
that the \raa's from the three collisions systems converge
for $\Npart>\sim$12, while a system ordering of
$R_{\rm dAu}$$>$$R_{\rm HeAu}$$>$$R_{\rm AuAu}$ is observed for
$\Npart<\sim$12. We haven't calculated the Integrated $R_{p{\rm A}}$
for $p$+Au collisions, but if we take the numbers at \pt=5\,\gevc,
the points are found to locate above $d$+Au points at $\Npart<\sim10$.

\begin{figure}[htbp]
\centering
\includegraphics[width=0.480\linewidth,clip]{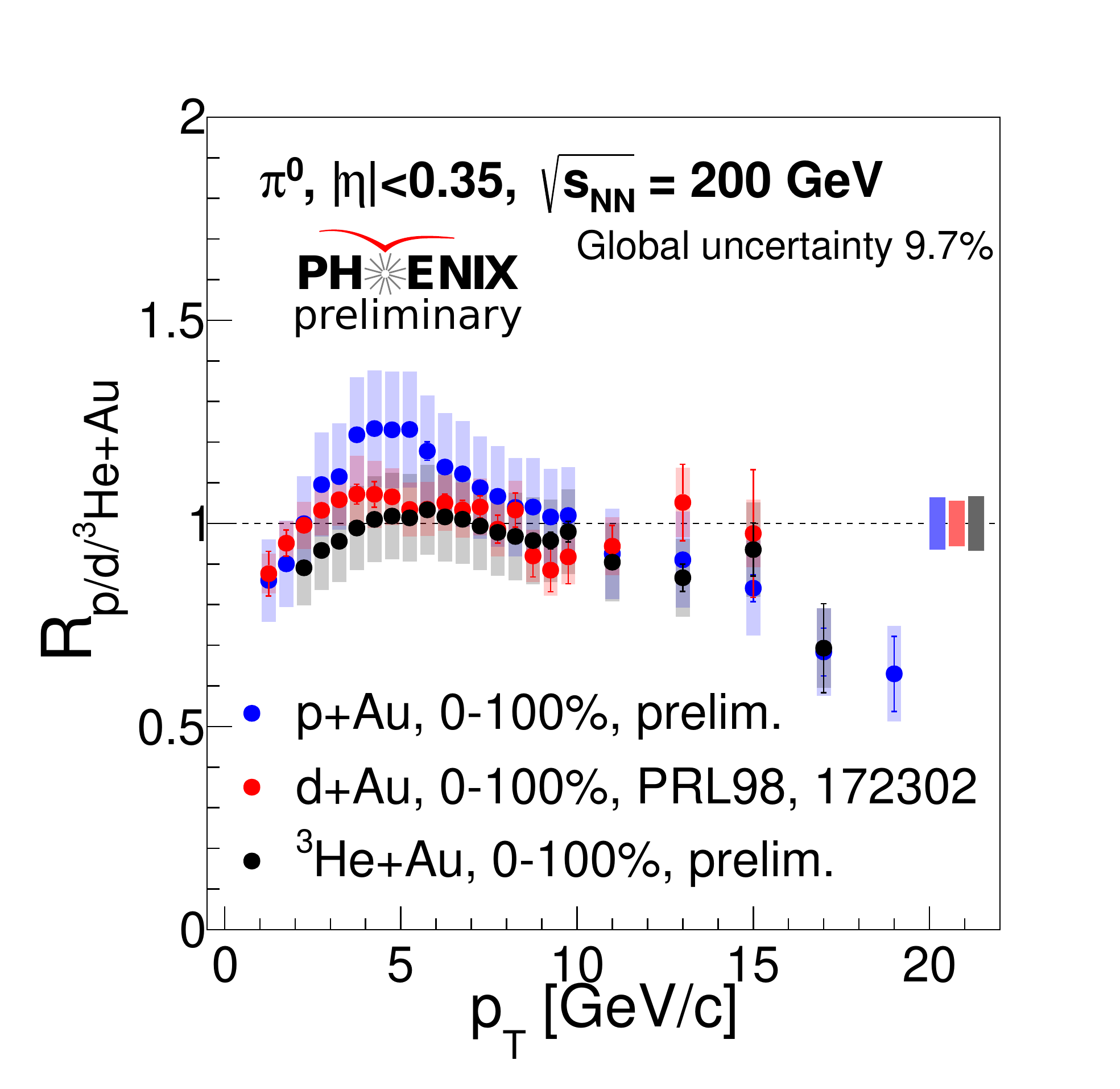}
\hspace{8mm}
\includegraphics[width=0.410\linewidth,clip]{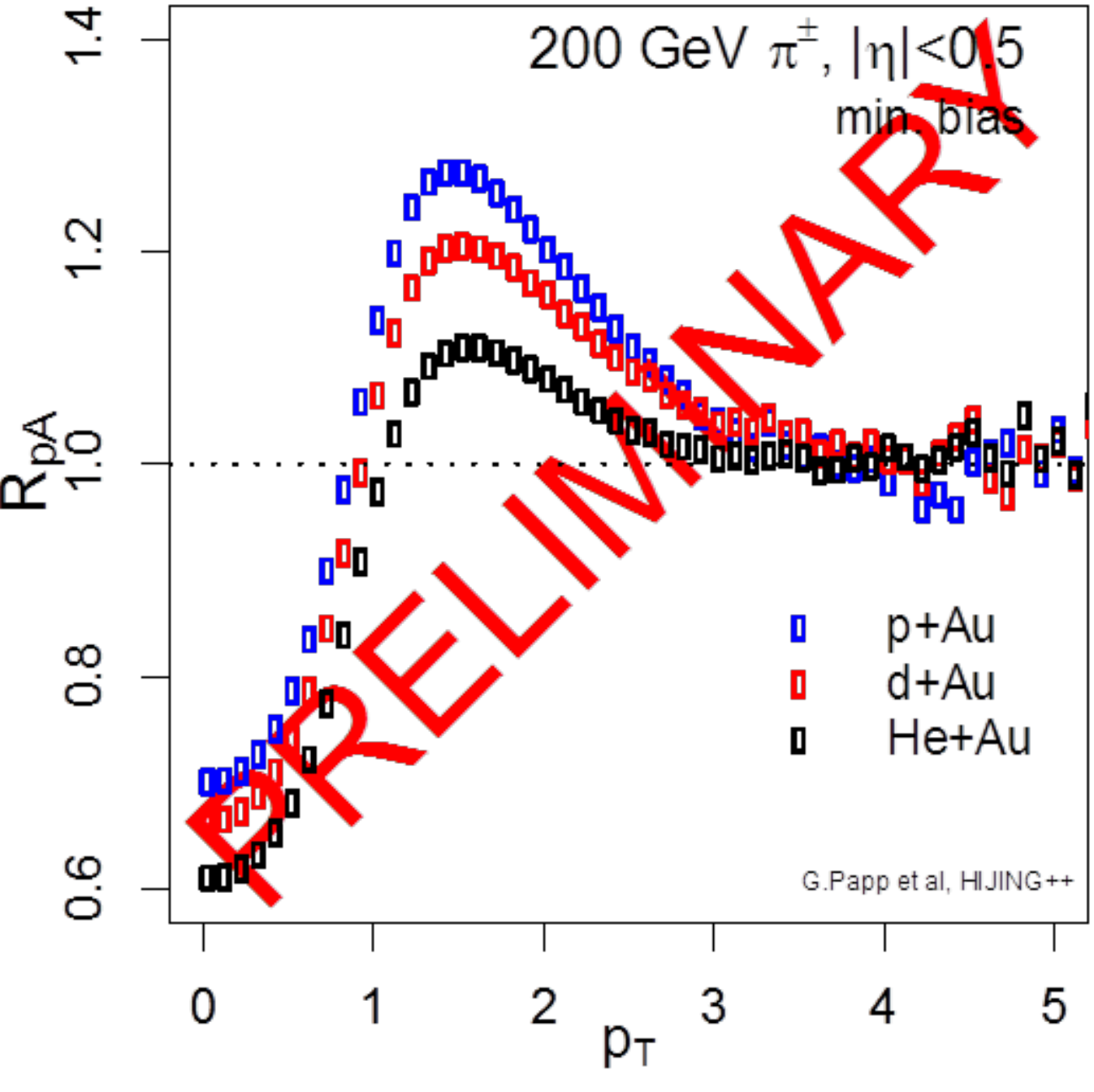}
\caption{(Left) $R_{p/d/{\rm HeA}}$ for minimum bias $p/d/^3$He+Au
collisions at \snn=200\,GeV. (Right) Calculation of the
$R_{p/d/{\rm HeA}}$ by HIJING++ model.}
\label{figHIJING}       
\end{figure}

We have made a comparison of the data with some models. A model
including cold energy loss only (e.g. ~\cite{Chien:2015vja})
was found not to reproduce the system ordering of the Cronin-peak
magnitude.
The HIJING++ model~\cite{Barnafoldi:2017jiz} was found to give
a reasonable description of the system ordering of the 
Cronin-peak as shown in Figure~\ref{figHIJING}.
It is, however, important to note that the peak positions are
different from what were observed in the data; the data have
the peaks at $\pt\sim5$\,\gevc, while the model has
the peaks at \pt$\sim$1.5\,\gevc.

\subsection{For(back)ward rapidity measurement}
\begin{figure}[htbp]
\centering
\includegraphics[width=0.420\linewidth,clip]{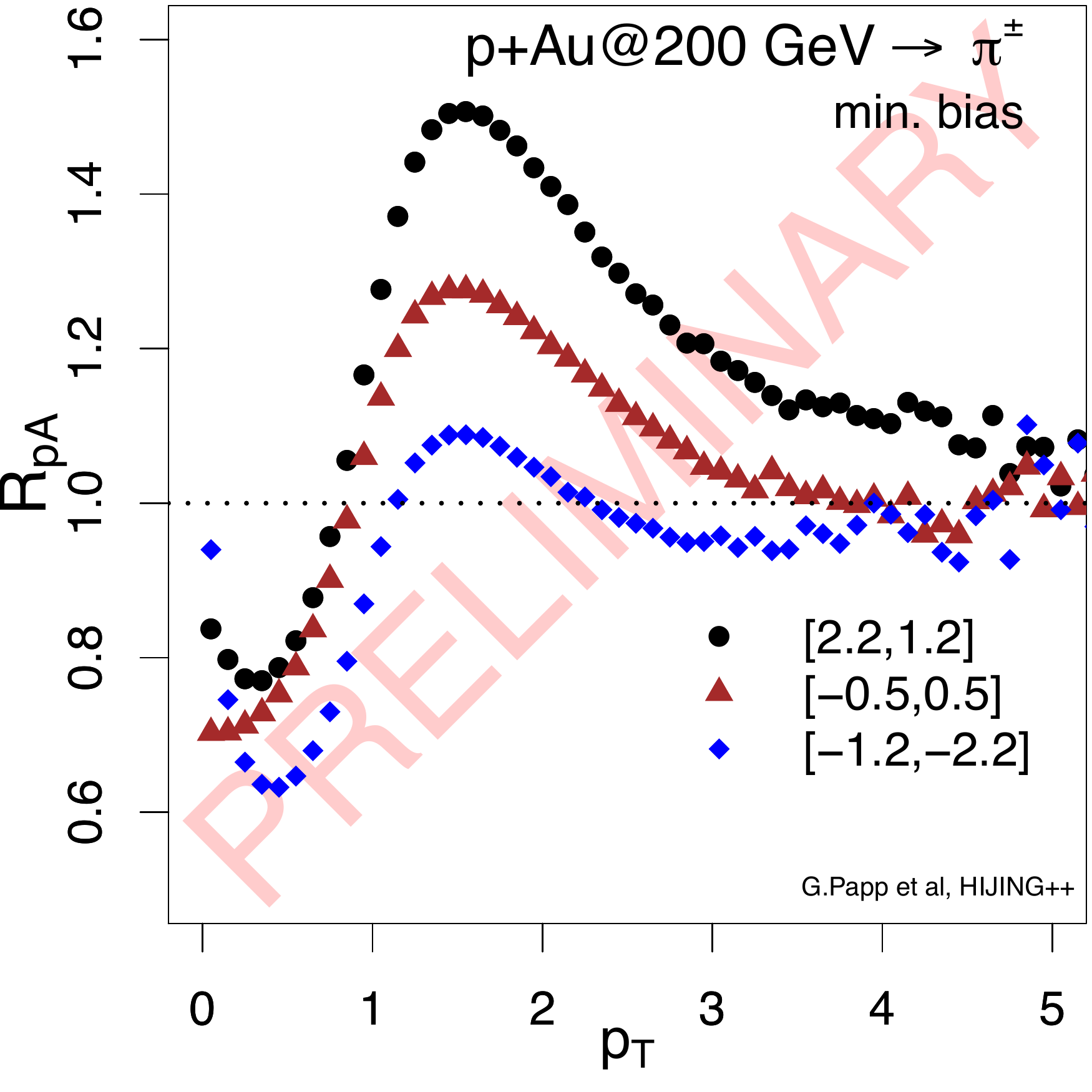}
\hspace{8mm}
\includegraphics[width=0.420\linewidth,clip]{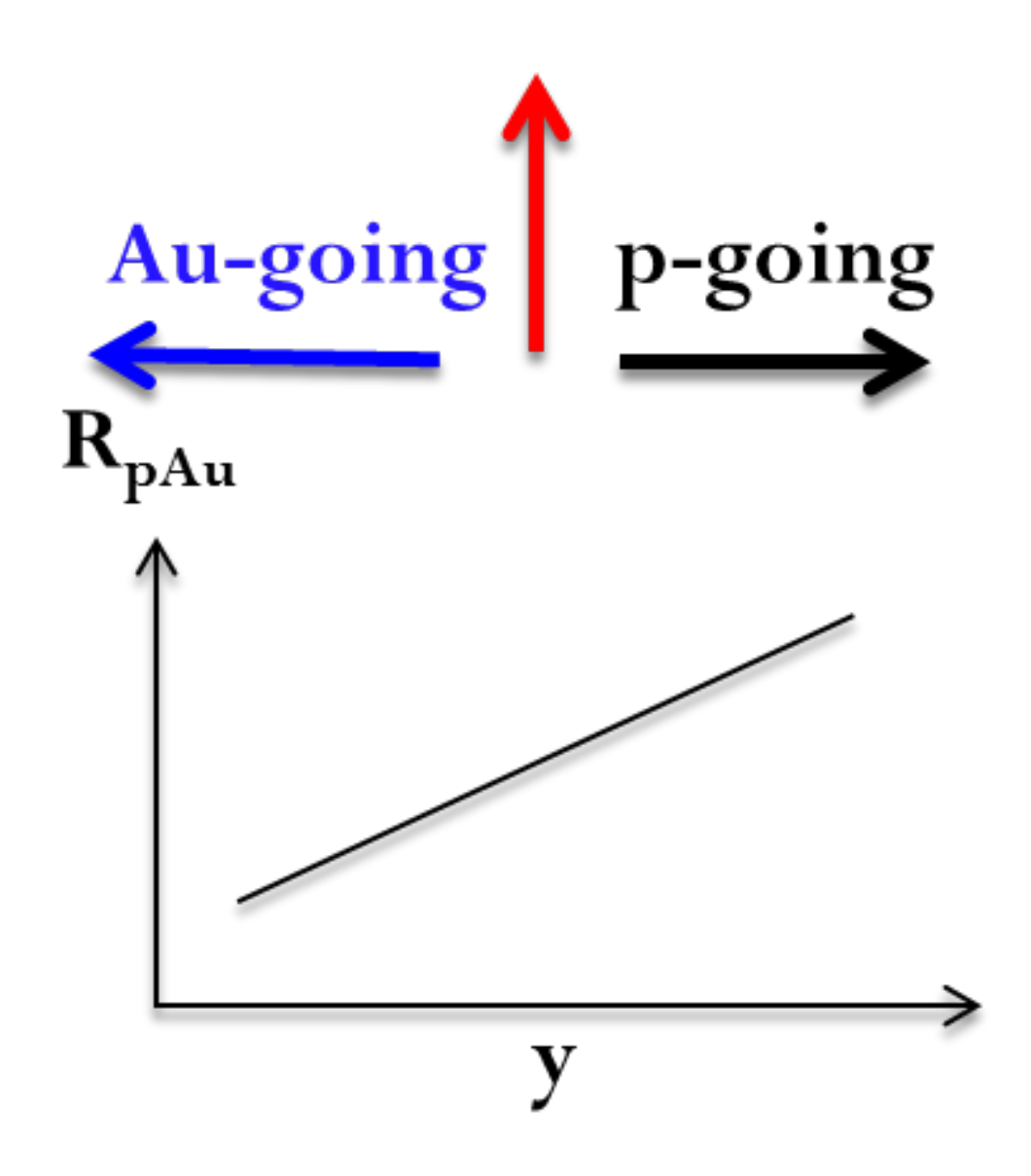}
\caption{(Left) HIJING++ prediction of $R_{p{\rm A}}$ for
pions for different rapidity regions in minimum bias $p$+Au collisions
at \snn=200\,GeV. (Right) Schematic drawing of the overall trend
of the $R_{p{\rm A}}$.}
\label{figHIJINGrap}       
\end{figure}

The HIJING++ model also gave a prediction of the rapidity
dependence of the magnitude of the Cronin-peak in $p$+Au
collisions as shown in Fig~\ref{figHIJINGrap}.
The prediction says that the $R_{p{\rm A}}$ increases as
going to $p$-going direction (positive rapidity), while
decreases as going to Au-going direction (negative rapidity).
This is mainly owing to the shadowing and anti-shadowing
effect in the parton distribution function in $p$- and
Au-going direction, respectively.

\begin{figure}[htbp]
\centering
\includegraphics[width=0.95\linewidth,clip]{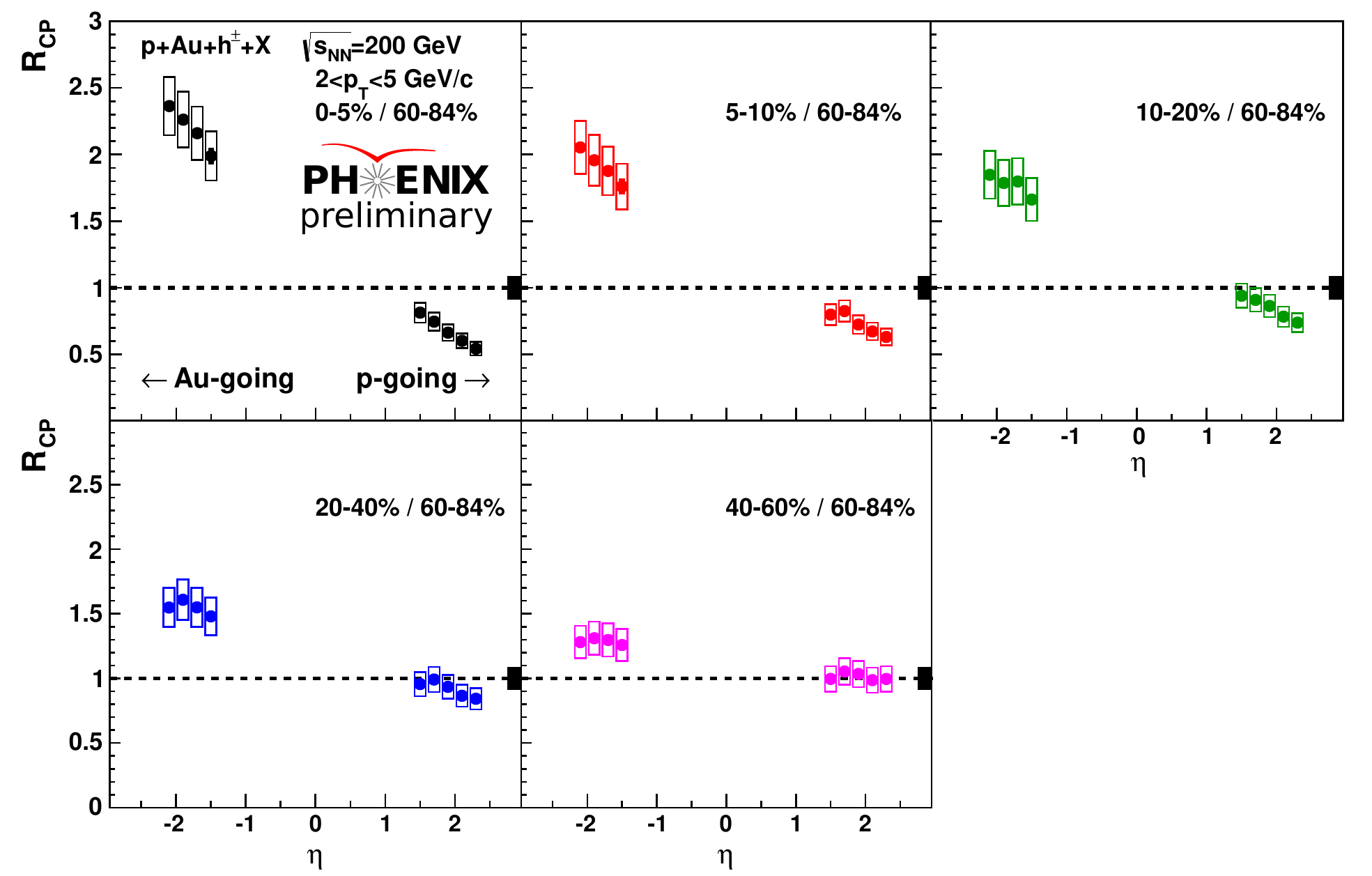}
\caption{Hadron \rcp in $p$+Au collisions at for(back)ward
rapidities as a function of centrality at \snn=200\,GeV.
The \pt range is 2$<\pt<$5\,\gevc.}
\label{figpAuRcpRap}       
\end{figure}

\begin{figure}[htbp]
\centering
\includegraphics[width=0.95\linewidth,clip]{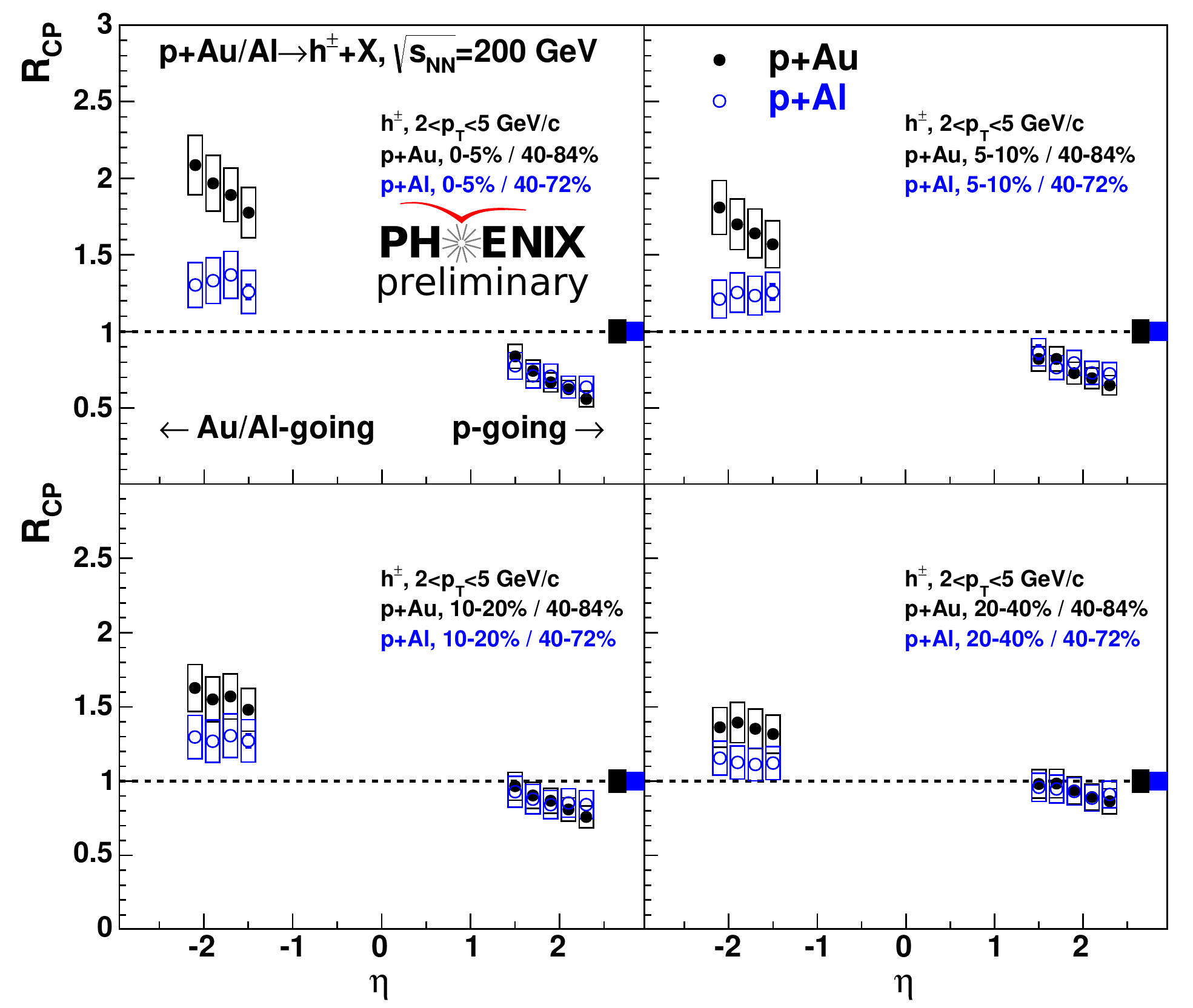}
\caption{Hadron \rcp in $p$+Al collisions at for(back)ward
rapidities as a function of centrality compared with those
in $p$+Au collisions at \snn=200\,GeV. The \pt range is
2$<\pt<$5\,\gevc.}
\label{figpAlRcpRap}       
\end{figure}

Figure~\ref{figpAuRcpRap} shows the centrality dependence of
the \rcp of hadrons for 2$<\pt<$5\,\gevc as a function of rapidity
in $p$+Au collisions at \snn=200\,GeV. 
The measurement was performed using the MuTr, FVTX and MuID
as described in the analysis section. We don't show the
measurement of midrapidity \piz in the plots, but we confirmed
that the \piz $R_{p{\rm A}}$ points are located right on the
linear interpolation of the for(back)ward measurements. It is
seen that the rapidity dependence of the \rcp
is apparently opposite to what was predicted by HINJING++. This
calls additional/another physics mechanism to be played.
Another observation is that the both enhancement and
suppression as a function rapidity is highly centrality
dependent, which is not expected from a parton distribution
function.
We have also measured the high \pt hadrons in for(back)ward
rapidity in $p$+Al collisions in RHIC Year-2015 as shown
in Figure~\ref{figpAlRcpRap}.
The result is compared with those in $p$+Au collisions. There
are interesting features seen in the result. First, although the
absolute magnitude are different, \rcp changes as a function of
centrality both in $p$+Au and $p$+Al collisions and both in
forward ($p$-going) and backward (Au/Al-going) rapidities;
in more central collisions, \rcp is more suppressed in forward
rapidity, and more enhanced in backward rapidity. The degree of
suppression at the forward rapidity are similar in $p$+Al/Au
collisions. On the contrary, the degree of enhancement at the
backward rapidity are very different as going to more central
collisions.

\subsection{Possible hot medium creation?}
\begin{figure}[htbp]
\centering
\includegraphics[width=1.0\linewidth,clip]{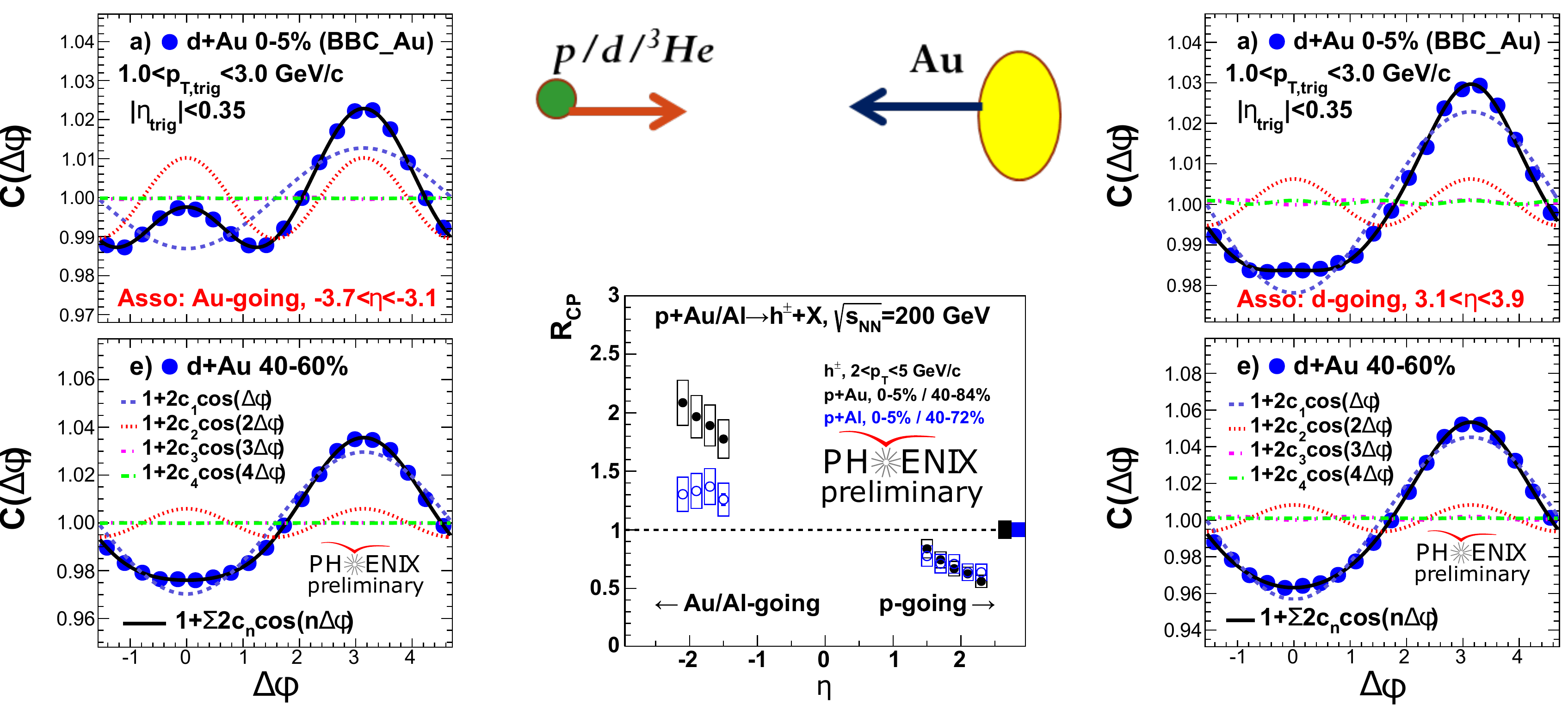}
\caption{(Left and Right) Two-particle correlation functions
measured in $d$+Au collisions at \snn=200\,GeV by PHENIX.
The trigger \pt is 1.0$<\pt<$3.0\,\gevc and the centralities are
0--5\,\% and 40--60\,\%. (Middle) the $R_{\rm cp}$ measurement
for 0--5\,\% centrality in $p$+Al/Au collisions at \snn=200\,GeV.
Note that Al ions go to the same directions as Au ions.}
\label{figFlowAndRcp}       
\end{figure}

We have measured a flow-like signal in $d$+Au collisions
through two-particle azimuthal angle correlation functions
with a wide rapidity gap ($|\Delta\eta|\sim3$) as mentioned
in the introduction section. We observed a quadrupole component
on top of a large dipole component when fitting the correlation
functions with Fourier series. We recently measured a similar
flow-like signal in $p$+Au collisions as well~\cite{Aidala:2016vgl}.
It is interesting to understand the different level of
enhancement in \rcp at backward rapidity in $p$+Al/Au collisions
in connection with the flow result.
Figure~\ref{figFlowAndRcp} shows the two-particle correlation
functions measured by PHENIX together with the hadron \rcp's
for 0--5\,\% centrality in $p$+Al/Au collisions.
The two-particle correlation functions shown at the left
panel are constructed by hadrons measured at CNT ($|\eta|<$0.35)
associated with the energy in a tower in the south MPC
(MPCS, -3.7$<\eta<$-3.1), which is located in Au-going side.
The functions shown at the right panel are constructed by
hadrons measured at CNT ($|\eta|<$0.35) associated with the
energy in a tower in the north MPC (MPCN, 3.1$<\eta<$3.9),
which is located in $p/d$-going side. The trigger \pt is
1.0$<\pt<$3.0\,\gevc and the centralities are 0--5\,\% and
40--60\,\%. On top of the dipole component in both cases,
the quadrupole component is
much prominent for Au-going side for 0--5\,\% centrality.
For the 40--60\,\% centrality, both $d$-going and Au-going
correlation functions look similar. The asymmetric feature of
the ridge/flow can be understood in the
context of the asymmetric pseudo-rapidity distribution of
produced particles in the asymmetric collision systems~\cite{Back:2004mr}.
In the middle panel, shown is the \rcp of the hadrons in forward
and backward rapidities. The rapidity dependent increase of
the \rcp is consistent with that of the ridge-like structure
in the two-particle correlation functions.
The possible scenario is that the enhancement of the
\rcp is partly due to the the increase of the
particle production by a possible hot medium created in the
collisions. Both the centrality dependent and system
size dependent increase of the \rcp in the backward rapidity
may be understood as follows; (1) the possible hot
medium creation is more prominent for more central collisions
and (2) for larger collision systems, and (3) the medium is
shifted towards Au-going direction for more central collisions.

\section{Summary}
High \pt hadrons in the small systems of $d/^3$He+Au and $p$+Al/Au
collisions have been measured at the midrapidity ($|\eta|<0.35$)
as well as
at the forward and backward rapidities ($1.2<|\eta|<2.4$) at
\snn=200\,GeV. A clear system and centrality ordering in the
$R_{p/d/{\rm HeA}}$ was observed for midrapidity \piz production.
Rapidity dependence of the Hadron \rcp in $p$+Au collisions show
an opposite trend to what HIJING++
predicted. The \rcp in $p$+Au collisions was found higher than
that of $p$+Al collisions at the backward rapidity, while the
\rcp at forward rapidity shows the similar behavior. This result
was compared with the two-particle long-range correlations
measured in $d$+Au collisions that are similar to $p$+Au collisions,
and found to be explained consistently with a possible hot medium
creation similar to A+A collisions.
%
\bibliography{SakaguchiProc}

\end{document}